\documentclass[conference]{IEEEtran}
\IEEEoverridecommandlockouts
\usepackage{cite}
\usepackage{amsmath,amssymb,amsfonts}
\usepackage{algorithmic}
\usepackage{graphicx}
\usepackage{textcomp}
\usepackage{xcolor}
\usepackage{stmaryrd}
\usepackage[english]{babel}
\newtheorem{theorem}{Theorem}
\newtheorem{Definition}{Definition}
\newtheorem{lemma}{Lemma}
\newtheorem{Proposition}{Proposition}

\usepackage{svg}
\def\BibTeX{{\rm B\kern-.05em{\sc i\kern-.025em b}\kern-.08em
    T\kern-.1667em\lower.7ex\hbox{E}\kern-.125emX}}
\begin{document}

\title{An Information-Spectrum Approach to Distributed Hypothesis Testing for General Sources}

\author{
 Ismaila Salihou Adamou$^{1}$, Elsa Dupraz$^{1}$, and Tad Matsumoto$^{1,2}$ \\
\small $^{1}$ IMT Atlantique, CNRS UMR 6285, Lab-STICC, Brest, France \\ $^{2}$ JAIST and University of
Oulu (Emeritus) \thanks{ This work has received a French government support granted to the Cominlabs excellence laboratory and managed by the National Research Agency in the ``Investing for the Future'' program under reference ANR-10-LABX-07-01.}
}



\maketitle 

\begin{abstract}
This paper investigates Distributed Hypothesis testing (DHT), in which a source $\mathbf{X}$ is encoded given that side information $\mathbf{Y}$ is available at the decoder only. Based on the received coded data, the receiver aims to decide on the two hypotheses $H_0$ or $H_1$ related to the joint distribution of $\mathbf{X}$  and $\mathbf{Y}$. While most existing contributions in the literature on DHT consider i.i.d. assumptions, this paper assumes more generic, non-i.i.d., non-stationary, and non-ergodic sources models. It relies on information-spectrum tools to provide general formulas on the achievable Type-II error exponent under a constraint on the Type-I error. The achievability proof is based on a quantize-and-binning scheme. It is shown that with the quantize-and-binning approach, the error exponent boils down to a trade-off between a binning error and a decision error, as already observed for the i.i.d. sources. The last part of the paper provides error exponents for particular source models, \emph{e.g.}, Gaussian, stationary, and ergodic models. 
\end{abstract}


\section{Introduction}
In distributed communication networks, data is gathered from various remote nodes and then sent to a server for further processing. Often, the primary objective of the server is not to reconstruct the data, but instead to make a decision based on the collected data. This type of setup is known as distributed hypothesis testing (DHT), and it was first investigated from an information-theoretic perspective in~\cite{Han1987,Ahlswede1986}.  

\begin{figure}[t!]
\centerline{\includegraphics[width=0.4\textwidth]{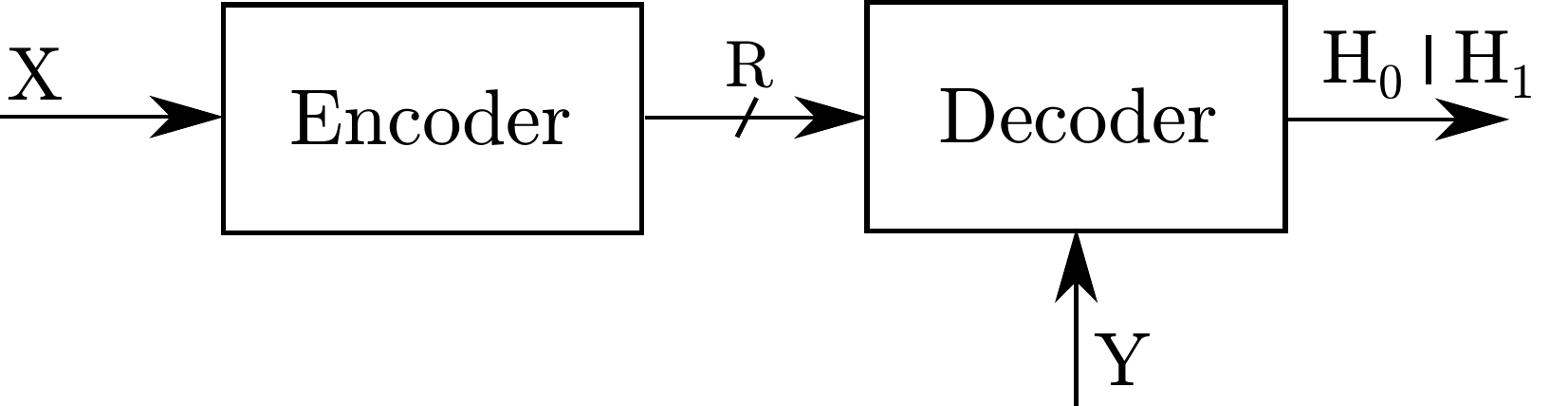}}
\caption{Distributed Hypothesis Testing coding scheme}
\label{system_model}
\end{figure}

In DHT, a source $\mathbf{X}$ is encoded using side information $\mathbf{Y}$ available only to the decoder, as shown in Figure~\ref{system_model}.
The receiver aims to make a decision between two hypothesis: $H_0$, where the joint probability distribution of $(\mathbf{X}, \mathbf{Y})$ is $P_{\mathbf{X}\mathbf{Y}}$, and $H_1$, where the joint distribution is $P_{\overline{\mathbf{X}}\overline{\mathbf{Y}}}$.
Hypothesis testing involves two types of errors, called the Type-I error and the Type-II error~\cite{lehmann2005testing}. 
The information-theoretic analysis of DHT aims to determine the achievable error exponent for the Type-II error, while keeping the Type-I error below a fixed threshold~\cite{Han1987,Ahlswede1986}. 

Previous contributions on DHT typically assume that the sources $\mathbf{X}$ and $\mathbf{Y}$ generate independent and identically distributed (i.i.d.) pairs of symbols $(X_t,Y_t)$~\cite{Katz2017,salehkalaibar2018distributed,Sreekumar2020,Rahman2012,Katz2015}. For example,~\cite{Rahman2012} and~\cite{Katz2015} provide the error exponent achieved by a quantize-and-binning scheme for i.i.d. sources.  
Some more complex source models have been investigated in~\cite{Rahman2009,Escamilla2020}, which assume that the sources $\mathbf{X}$ and $\mathbf{Y}$ generate pairs of Gaussian vectors $(\mathbf{X}_t^M, \mathbf{Y}_t^N)$ with auto-correlations embedded in each vector $\mathbf{X}_t^M$ and $\mathbf{Y}_t^N$, as well as cross-correlation between them.
However, the models of~\cite{Rahman2009,Escamilla2020} are block-i.i.d. in the sense that the successive pairs $(\mathbf{X}_t^M, \mathbf{Y}_t^N)$ are assumed to be i.i.d. with $t$.

However, i.i.d. and block-i.i.d. models are often inadequate for capturing the statistics of signals like time series or videos, which cannot be decomposed into fixed-length independent blocks and are frequently non-stationary and/or non-ergodic. As a result, the objective of this paper is to consider a more general source model that is non-i.i.d. and can account for non-stationary and non-ergodic signals, while still encompassing the previous models as particular instances.
To investigate DHT under these conditions, we utilize information spectrum tools, which were first introduced in~\cite{Han1998} and generally provide information theory results that are applicable to a broad range of source models. It should be noted that information spectrum has been previously used for hypothesis testing in~\cite{Han2000}, but only for the encoding of a source $\mathbf{X}$ alone, without the use of side information $\mathbf{Y}$.

In this paper, we investigate DHT using general source models for $\mathbf{X}$ and $\mathbf{Y}$ and provide an achievability scheme that yields a general expression for the Type-II error exponent. Our approach to the achievability scheme builds upon the quantize-and-binning techniques presented in~\cite{Katz2015}, while taking into account the use of side information for more complex source models. As in~\cite{Katz2015}, the resulting error exponent consists of two terms: one for the binning error and the other for the decision error. We then specialize our error-exponent to specific source models of interest, including (i)~i.i.d. sources, for which we recover the error exponent reported in~\cite{Katz2015}; (ii)~non-i.i.d. stationary and ergodic sources in general; and (iii)~non-i.i.d. Gaussian stationary and ergodic sources.

The outline of the paper is as follows. Section~\ref{Problem_statement} describes the general sources model and restates the DHT problem. Section~\ref{main_theoritical_result} provides the achievable error exponent for general sources, and Section~\ref{proof} derives the proof. Section \ref{examples} considers some examples of source models.

\section{Problem statement} \label{Problem_statement}
In the DHT problem shown in Fig.\ref{system_model}, the encoder observes a source sequence $\mathbf{X}$, and the decoder receives a coded version of $\mathbf{X}$ as well as a side information sequence $\mathbf{Y}$, where $\mathbf{X}$ and $\mathbf{Y}$ are correlated. In what follows, $\llbracket 1,n\rrbracket$ denotes the set of integers between $1$ and $n$. 
\subsection{General Sources} \label{general_sources}
We consider that the sequences $\mathbf{X}$ and $\mathbf{Y}$ are produced from two general sources which are not necessarily i.i.d., and not even stationary or ergodic. As in~\cite{Han2000}, we define general sources $\mathbf{X}$ and $\mathbf{Y}$ as two infinite sequences : 
 \begin{align}
 \mathbf{X} & =\left\{\mathbf{X}^n=\left(X_1, X_2, \cdots, X_n\right)\right\}_{n=1}^{\infty},
     \notag \\
\mathbf{Y}& =\left\{\mathbf{Y}^n=\left(Y_1, Y_2, \cdots, Y_n\right)\right\}_{n=1}^{\infty}
     \label{general_Y}
 \end{align}
 of $n$-dimensional random variables $\mathbf{X}^n, \mathbf{Y}^n$, respectively. Each component random variable $X_i, Y_i$, $i \in \llbracket 1,n\rrbracket$, takes values in a finite source alphabet $\mathcal{X}, \mathcal{Y}$, respectively. 
In what follows, $P_{\mathbf{X}^n}$ is the probability distribution of the length-n vector $\mathbf{X}^n$, and $P_{\mathbf{X}} = \left\{ P_{\mathbf{X}^n}\right\}_{n=1}^{\infty}$ is the collection of all probability distributions $P_{\mathbf{X}^n}$. The same holds for the source $\mathbf{Y}$.
 
 We now describe two particular cases of \eqref{general_Y}. The first one consists of a scalar i.i.d. model in which the sequences $\mathbf{X}^n$ and $\mathbf{Y}^n$ come from two i.i.d. sources, \emph{i.e.}, the successive pairs of symbols $\left(X_n, Y_n\right)$ are independent and distributed according to the same joint distribution $P_{XY}$. This model was considered for DHT in~\cite{Rahman2012,Katz2015}. The second case still relies on an i.i.d. model but for source vectors. In this case, the source sequences $\mathbf{X}^n$ and $\mathbf{Y}^n$ are defined as 
 \begin{equation}
\mathbf{X}^n = \left\{\mathbf{X}_{t}^{M}\right\}_{t=1}^n, ~ \mathbf{Y}^n= \left\{\mathbf{Y}_{t}^{N}\right\}_{t=1}^n,
     \label{vector_Y}
 \end{equation}
 where $\{\mathbf{X}_t^M\}_{t=1}^n$ and $\{\mathbf{Y}_t^N\}_{t=1}^n $ are sequences of i.i.d. M-dimensional and N-dimensional random vectors, respectively. This means that successive pairs of $\left(\mathbf{X}_t^M,\mathbf{Y}_t^N\right)$ are independent and distributed according to the same joint distribution $P_{\mathbf{X}^M\mathbf{Y}^N}$. The i.i.d. property of the successive M-length and N-length vectors simplifies the DHT analysis by allowing for an orthogonal transform to be applied onto the successive independent blocks $\mathbf{X}_t^M$ and $\mathbf{Y}_t^N$~\cite{Rahman2009,Escamilla2020}. Our model described in~\eqref{general_Y} is more general since it considers infinite sequences without the i.i.d. assumption. 
\subsection{Distributed Hypothesis Testing} 
In what follows, we consider that the joint distribution of the sequence pair $\left\{\left(\mathbf{X}^{n},\mathbf{Y}^{n}\right)\right\}_{n=1}^{\infty}$ depends on the underlying hypotheses $H_0$ and $H_1$ defined as
\begin{equation}
    H_0  : \left(\mathbf{X}^{n},\mathbf{Y}^{n}\right) \sim P_{\mathbf{X}^{n}\mathbf{Y}^{n}}, 
    \label{H_0}
\end{equation}
\begin{equation}
    H_{1}  : \left(\mathbf{X}^{n},\mathbf{Y}^{n}\right) \sim P_{\overline{\mathbf{X}}^{n}\overline{\mathbf{Y}}^{n}}.
    \label{H_1}
\end{equation}
where the marginal probability distributions $P_{\mathbf{X}^{n}}$ and $P_{\mathbf{Y}^{n}}$ do not depend on the hypothesis. 

We consider the following usual coding scheme defined in the literature on DHT \cite{Han1987,Katz2015}. 

\begin{Definition} The encoding function $f^{(n)}$ and decoding function $g^{(n)}$ are defined as
\begin{align}
f^{(n)} &: \mathcal{X}^{n} \longrightarrow \mathcal{M}_{n}= \llbracket 1,M_2\rrbracket, 
    \label{enco} \\
g^{(n)} &: \mathcal{M}_{n}\times\mathcal{Y}^{n} \longrightarrow \mathcal{H}=\{H_{0}, H_{1}\},
    \label{dec}
\end{align}
such that
\begin{equation}
\limsup_{n \rightarrow \infty} \frac{1}{n} \log  \text{M}_{2} \leq \text{R},
    \label{rate_constraint}
\end{equation}
where $R$ is the rate and $\text{M}_{2}$ is the cardinality of the alphabet set $\mathcal{M}_{n}$. 
\label{defintion_f_g}
\end{Definition}
\begin{Definition} The Type-I and Type-II error probabilities $\alpha_n$ and $\beta_n$ are defined as  
\begin{align}
\alpha_n &=\mathbb{P}\left[g^{(n)}\left(f^{(n)}\left(\mathbf{X}^{n}\right),\mathbf{Y}^{n}\right)=H_{1}\mid H_{0} \ \text{is true}\right], \label{alpha}\\
\beta_n &=\mathbb{P}\left[g^{(n)}\left(f^{(n)}\left(\mathbf{X}^{n}\right),\mathbf{Y}^{n}\right)=H_{0}\mid H_{1} \ \text{is true}\right] .
\label{beta}
\end{align}
\end{Definition}
\begin{Definition}
The Type-II error exponent $\theta$ is said to be achievable for a given rate $R$, if for large blocklength $n$, there exists encoding and decoding functions  $\left(f^{(n)}, g^{(n)}\right)$ such that the Type-I and Type-II error probabilities $\alpha_n$ and $\beta_n$ satisfy  
\begin{equation}
    \alpha_n \leq \epsilon ,
    \label{alpha_const}
\end{equation}
and
\begin{equation}
    \limsup_{n \rightarrow \infty} \frac{1}{n} \log \frac{1}{\beta_{n}} \geq \theta 
    \label{betha_const}
\end{equation}
for any $\epsilon>0$.
\label{defintion_alpha_betha}
\end{Definition}

In the following, we aim to determine the achievable Type-II error exponent $\theta$ for general sources.

\section{Main result: Error exponent} \label{main_theoritical_result}
\subsection{Definitions}
We first provide some definitions which will be useful to express our main result. The  $\limsup$ and $\liminf$ in probability of a sequence $\left\{Z_n\right\}_{n=1}^{\infty}$ are, respectively, defined as \cite{Han1998}
\begin{align}
    \mathrm{p}-\limsup _{n \rightarrow \infty} Z_n & =\inf \left\{\alpha \mid \lim _{n \rightarrow+\infty} \mathbb{P}\left(Z_n>\alpha\right)=0\right\} ,
    \label{p_limsup} 
    \\
    \mathrm{p}-\liminf _{n \rightarrow \infty} Z_n & =\sup \left\{\alpha \mid \lim _{n \rightarrow+\infty} \mathbb{P}\left(Z_n<\alpha\right)=0\right\}
    \label{p_liminf}.
\end{align}
 The spectral sup-mutual information $\overline{I}(\mathbf{X} ; \mathbf{U})$, the spectral inf-mutual information $\underline{I}(\mathbf{U} ; \mathbf{Y})$, the spectral inf-divergence rate $\underline{D}\left(P_{\mathbf{U}\mathbf{Y}} \| P_{\overline{\mathbf{U}}\overline{\mathbf{Y}}}\right)$, and the spectral sup-divergence rate $\overline{D}\left(P_{\mathbf{U}\mathbf{Y}} \| P_{\overline{\mathbf{U}}\overline{\mathbf{Y}}}\right)$ are, respectively, defined as~\cite{Han1998} 
 \begin{align}
    & \bar{I}(\mathbf{X} ; \mathbf{U}) =\mathrm{p}-\limsup _{n \rightarrow \infty} \frac{1}{n} \log \frac{P_{\mathbf{U}^n \mid \mathbf{X}^n}\left(\mathbf{U}^n \mid \mathbf{X}^n\right)}{P_{\mathbf{U}^n}\left(\mathbf{U}^n\right)},
    \label{sup_mutual} \\
  &  \underline{I}(\mathbf{U} ; \mathbf{Y}) =\mathrm{p}-\liminf _{n \rightarrow \infty} \frac{1}{n} \log \frac{P_{\mathbf{U}^n \mid \mathbf{Y}^n}\left(\mathbf{U}^n \mid \mathbf{Y}^n\right)}{P_{\mathbf{U}^n}\left(\mathbf{U}^n\right)},
    \label{inf_mutual}
\\
   & \underline{D}\left(P_{\mathbf{U}\mathbf{Y}} \| P_{\overline{\mathbf{U}}\overline{\mathbf{Y}}}\right)  = \mathrm{p}-\liminf _{n \rightarrow \infty}\frac{1}{n} \log \frac{P_{\mathbf{U}^n \mathbf{Y}^n}\left(\mathbf{U}^n , \mathbf{Y}^n\right)}{P_{\overline{\mathbf{U}}^n \overline{\mathbf{Y}}^n}\left(\mathbf{U}^n , \mathbf{Y}^n\right)},
   \label{inf-div}
\\
   & \overline{D}\left(P_{\mathbf{U}\mathbf{Y}} \| P_{\overline{\mathbf{U}}\overline{\mathbf{Y}}}\right) = \mathrm{p}-\limsup _{n \rightarrow \infty}\frac{1}{n} \log \frac{P_{\mathbf{U}^n \mathbf{Y}^n}\left(\mathbf{U}^n , \mathbf{Y}^n\right)}{P_{\overline{\mathbf{U}}^n \overline{\mathbf{Y}}^n}\left(\mathbf{U}^n , \mathbf{Y}^n\right)}.
   \label{sup-div}
\end{align}
\subsection{Achievable error-exponent for general sources}
\begin{theorem} For the coding scheme of Definition~\ref{defintion_f_g}, the following error exponent $\theta$ is achievable for general sources defined by~\eqref{general_Y}: 
\begin{align}\notag
    \theta \leq \min & \left\{ r-\left(\overline{I}(\mathbf{X} ; \mathbf{U})-\underline{I}(\mathbf{U} ; \mathbf{Y})\right), \right. \\
    & \left. \underline{D}\left(P_{\mathbf{U}\mathbf{Y}} \| P_{\overline{\mathbf{U}}\overline{\mathbf{Y}}}\right) + \left( \underline{I}(\mathbf{X} ; \mathbf{U}) - \overline{I}(\mathbf{X} ; \mathbf{U}) \right)\right\},
    \label{th1}
\end{align}
where $\mathbf{U}$ is an auxiliary random variable with same conditional distribution $P_{\mathbf{U}|\mathbf{X}} = P_{\mathbf{\overline{U}}|\mathbf{\overline{X}}} $ under $H_0$ and $H_1$ and such that the Markov chain $\mathbf{U} \rightarrow \mathbf{X} \rightarrow \mathbf{Y}$ is satisfied under both $H_0$ and $H_1$. In addition, $P_{\mathbf{U}\mathbf{Y}}$, and $P_{\overline{\mathbf{U}}\overline{\mathbf{Y}}}$ are the joint distributions of $\left(\mathbf{U}^n, \mathbf{Y}^n \right)$ under $H_0$ and $H_1$, respectively, and $r \leq \text{R}$. 
\label{theorem_1}
\end{theorem}

The error exponent \eqref{th1} is achieved by a quantize-and-binning strategy in which the decoder works in two steps. First, it looks for a sequence in the bin according to the joint distribution $P_{\mathbf{U}^n\mathbf{Y}^n}$ under $H_0$. Then, it declares $H_0$ by checking that if the sequence extracted from the bin belongs to a certain acceptance region $\mathcal{A}_n$ defined in \eqref{A_n}; if otherwise, it declares $H_1$. The binning strategy introduces a new type of error event which does not appear in the DHT scheme without binning for general sources of~\cite{Han2000}. Therefore, the error exponent in \eqref{th1} is the result of a trade-off between the binning error and the decision error, as in the i.i.d. case~\cite{Katz2015,Rahman2012}.
In addition, the decision error, \emph{e.g.}, the second term in~\eqref{th1}, not only contains a divergence term that appears in~\cite{Katz2015,Rahman2012} and related works, but also the difference $\underline{I}(\mathbf{X} ; \mathbf{U}) - \overline{I}(\mathbf{X} ; \mathbf{U})$ between the spectral inf-mutual information and the spectral sup-mutual information of $\mathbf{X}$ and $\mathbf{U}$. Especially, if the term  $\frac{1}{n} \log \frac{P_{\mathbf{U}^n \mid \mathbf{X}^n}\left(\mathbf{U}^n \mid \mathbf{X}^n\right)}{P_{\mathbf{U}^n}\left(\mathbf{U}^n\right)}$ does not converge in probability, then the two mutual information terms differ, inducing a penalty in the error exponent. For stationary and ergodic sources, this term converges and there is no penalty.  


\section{Proof of Theorem \ref{theorem_1}} \label{proof}
We first restate the following lemma from~\cite{iwata2002information}, which will be useful in the proof.
\begin{lemma}[\hspace{-0.01cm}\cite{iwata2002information}] Let $\mathbf{Z}^n, \mathbf{X}^n$,  $\mathbf{U}^n$, be random sequences which take values in finite sets $\mathcal{Z}^n, \mathcal{X}^n$, $\mathcal{U}^n$, respectively, and satisfy the Markov condition $\mathbf{U}^n \rightarrow \mathbf{X}^n \rightarrow \mathbf{Z}^n $.
Let $\left\{\Psi_n\right\}_{n=1}^{\infty}$ be a sequence of mappings such that $\Psi_n: \mathcal{Z}^n \times \mathcal{U}^n \rightarrow\{0,1\}
$, and
\begin{equation}
 \lim _{n \rightarrow \infty} \mathbb{P}\left(\Psi_n(\mathbf{Z}^n, \mathbf{U}^n)=1\right)=0 .
\end{equation}
Then, $\forall \varepsilon>0$, there exists a sequence $\left\{f_{n}\right\}_{n=1}^{\infty}$ of mappings $f_{n}: \mathcal{X}^n \rightarrow\left\{\mathbf{u}_i^n\right\}_{i=1}^{\text{M}} \subset \mathcal{U}^n$ such that $\text{M}=\lceil e^{n(\overline{I}(\mathbf{U} ; \mathbf{X})+\varepsilon)}\rceil$ and
\begin{equation}
\lim _{n \rightarrow \infty} \mathbb{P}\left(\Psi_n(\mathbf{Z}^n, f_{n}(\mathbf{X}^n))=1\right)=0 .
\end{equation}
\end{lemma}
\subsection{Coding scheme}
\textit{Random codebook generation:} Generate $\text{M}{_1}= e^{n\overline{r}_0}$ sequences $\mathbf{u}_i^{n}$ randomly according to a fixed distribution $P_{\mathbf{U}^n \mid \mathbf{X}^n}$. Assign randomly each $\mathbf{u}_i^{n}$ to one of $\text{M}_{2}=e^{nr}$ bins according to a uniform distribution over $ \llbracket 1,\text{M}_{2} \rrbracket$. Let $\boldsymbol{B}(\mathbf{u}_i^{n}) \in \llbracket 1,\text{M}_{2} \rrbracket$ denote the index of the bin to which $\mathbf{u}_i^{n}$ belongs to.

\textit{Encoder :} Given the sequence $\mathbf{x^{n}}$, the encoder uses a pre-defined mapping $f_n:\mathcal{X}^n \rightarrow \left\{\mathbf{u}_i^n\right\}_{i=1}^{\text{M}_1}$ to output a certain sequence $ \mathbf{u}_i^n =f_{n}(\mathbf{x}^n)$ and checks if the condition $(\mathbf{x^{n}}, \mathbf{u_i^{n}}) \in  T_{n}^{(1)}$ is satisfied, where
\begin{align}\label{T_1}
   & T_{n}^{(1)} = \\ & \left\lbrace\left(\mathbf{x}^n, \mathbf{u}^n\right) \text { s.t. } \underline{r}_{0}  -\epsilon < \frac{1}{n} \log \frac{P_{\mathbf{U}^n \mid \mathbf{X}^n}\left(\mathbf{u}^n \mid \mathbf{x}^n\right)}{P_{\mathbf{U}^n}\left(\mathbf{u}^n\right)} < \notag  \overline{r}_0+\epsilon\right\rbrace.
\end{align}
 If such a sequence is found, the encoder sends the bin index $\boldsymbol{B}(\mathbf{u}_i^{n})$. 
 Otherwise, it sends an error message.
 \hspace{0.1cm}
 
\textit{Decoder :} The decoder first looks for a sequence in the bin according to the joint distribution $P_{\mathbf{U}^n\mathbf{Y}^n}$ under $H_0$. Given the received bin index and the side information $\mathbf{y}^{n}$, going over the sequences $\mathbf{u}^n$ in the bin one by one, the decoder checks whether $(\mathbf{y^{n}}, \mathbf{u^{n}}) \in T_{n}^{(2)}$ with
\begin{equation}
    T_{n}^{(2)} = \left\lbrace\left(\mathbf{y}^n, \mathbf{u}^n\right) \text { s.t. } \frac{1}{n} \log \frac{P_{\mathbf{U}^n \mid \mathbf{Y}^n}\left(\mathbf{u}^n \mid \mathbf{y}^n\right)}{P_{\mathbf{U}^n}\left(\mathbf{u}^n\right)}>r^{'}-\epsilon\right\} .
    \label{T-2}
\end{equation}
The decoder declares $H_{1}$ if no such sequence is found in the bin or if it receives an error message from the encoder. Otherwise, it declares $H_0$ if the sequence $\mathbf{u}^n$ extracted from the bin belongs to the acceptance region $\mathcal{A}_n$ defined as
\begin{equation}
         \mathcal{A}_n =\left\lbrace\left(\mathbf{y}^n, \mathbf{u}^n\right) \text{ s.t.}  \ \   \frac{1}{n}  \log\frac{P_{\mathbf{U}^n \mathbf{Y}^n}\left(\mathbf{u}^n , \mathbf{y}^n\right)}{P_{\overline{\mathbf{U}}^n \overline{\mathbf{Y}}^n}\left(\mathbf{u}^n , \mathbf{y}^n\right)} > S - \epsilon \right\rbrace
,
\label{A_n}
\end{equation}
where $S$ is the decision threshold; if otherwise, it declares $H_1$.
\hspace{0.1cm}
%
\subsection{Error probability analysis}
\textit{Type-I error\ $\alpha_{n}$ :} The error events with which the decoder declares $H_1$ under $H_0$  are as follows: 
\begin{align}
   E_{11} & =\left\lbrace\nexists \mathbf{u}^{n} \text { s.t. }\left(\mathbf{X}^n, \mathbf{u}^n\right) \in T_{n}^{(1)}, \left(\mathbf{Y}^n, \mathbf{u}^n\right) \in T_{n}^{(2)},  \right. \notag \\
     & \left. \left(\mathbf{Y}^n, \mathbf{u}^n\right) \in \mathcal{A}_n  \right\rbrace, 
    \label{E_11} \\
      E_{12}& =\left\lbrace \exists \mathbf{u'}^{n} \neq \mathbf{u}^n \text{ s.t. } \boldsymbol{B}(\mathbf{u'}^{n})=\boldsymbol{B}(\mathbf{u}^{n}) , \left(\mathbf{Y}^n, \mathbf{u'}^n\right) \in T_{n}^{(2)}, \right. \notag \\
     & \text{but}\left. \ (\mathbf{Y}^{n},\mathbf{u'}^{n})\notin \mathcal{A}_n\right\rbrace.
     \label{E_12}
\end{align}
The first event $E_{11}$ is when there is an error either in the encoding, during debinning, or when taking the decision. The second event $E_{12}$ corresponds to a debinning error, where a wrong sequence is extracted from the bin. By the union-bound, the Type-I error probability $\alpha_n$ can be upper bounded as
\begin{equation}
    \alpha_n \leq \mathbb{P}\left(E_{11}\right)+\mathbb{P}\left(E_{12}\right) .
    \label{alpha}
\end{equation}
Regarding the first error event, for $\underline{r}_0 = \underline{I}(\mathbf{X} ; \mathbf{U})$, $\overline{r}_0 =  \overline{I}(\mathbf{X} ; \mathbf{U})$, and from the definitions of $\underline{I}(\mathbf{X} ; \mathbf{U})$ and $\overline{I}(\mathbf{X} ; \mathbf{U})$ in~\eqref{sup_mutual} and~\eqref{inf_mutual}, we have $$\lim_{n\rightarrow \infty} \mathbb{P}\left(\left(\mathbf{X}^n, \mathbf{U}^n\right) \notin T_{n}^{(1)}\right) = 0.$$ 
In addition, according to the definition of $\underline{I}(\mathbf{Y} ; \mathbf{U})$ in \eqref{inf_mutual}, and setting $r^{'}=\underline{I}(\mathbf{Y} ; \mathbf{U})$, we also have  \begin{equation}\label{eq:Pe1An}\lim_{n\rightarrow \infty} \mathbb{P}\left(\left(\mathbf{Y}^n, \mathbf{U}^n\right) \notin T_{n}^{(2)}\right) = 0.\end{equation} Finally, when $S=\underline{D}\left(P_{\mathbf{U}\mathbf{Y}} \| P_{\overline{\mathbf{U}}\overline{\mathbf{Y}}}\right)$ and from the definition of $\underline{D}\left(P_{\mathbf{U}\mathbf{Y}} \| P_{\overline{\mathbf{U}}\overline{\mathbf{Y}}}\right)$, we have $$ \lim_{n\rightarrow \infty} \mathbb{P}\left(\left(\mathbf{Y}^n, \mathbf{U}^n\right) \notin \mathcal{A}_n\right) = 0.$$ Thus, by defining
\begin{align}
    & \Psi_n(\mathbf{x}^n,\mathbf{y}^n,\mathbf{u}^n) =  \\ \notag & \left\{\begin{array}{ll}
0, & \text { if } (\mathbf{x}^{n},  \mathbf{u}^{n}) \in T_n^{(1)}, (\mathbf{y}^{n},  \mathbf{u}^{n}) \in T_n^{(2)} \text{ and } \\&(\mathbf{y}^{n},  \mathbf{u}^{n}) \in \mathcal{A}_n, \\
1, & \text{otherwise.}
\end{array}\right.
\end{align}
we get that $\mathbb{P}(\Psi_n(\mathbf{X}^n,\mathbf{Y}^n,\mathbf{U}^n) = 1) \rightarrow 0$ as $n \rightarrow \infty$. 
Then, given that $\mathbf{U}^n \rightarrow \mathbf{X}^n \rightarrow \mathbf{Y}^n$ forms a Markov chain, applying Lemma $1$ allows to show that there exists a sequence of functions $f_{n}$ such that $\mathbb{P}(E_{11}) \rightarrow 0$ as $n\rightarrow \infty$. 

Then, the error probability $\mathbb{P}\left(E_{12}\right)$ can be expressed as
 \begin{align}
     \mathbb{P}\left(E_{12}\right) & \leq 
        \sum_{ \mathbf{y}^{n}} P_{\mathbf{Y}^{n}}\left(\mathbf{y}^{n}\right) \hspace{-0.8cm}\sum_{\substack{\mathbf{u'}^n:\mathbf{u'}^n \neq \mathbf{u}^n \\ (\mathbf{y}^n,\mathbf{u'}^n) \in \mathcal{T}_n^{(2)}\cap{\overline{\mathcal{A}_n}}}} \hspace{-0.3cm} \mathbb{P} \left(\boldsymbol{B}(\mathbf{u'}^{n})=\boldsymbol{B}(\mathbf{u}^{n})\right) \notag \\
       & \leq \sum_{ \mathbf{y}^{n}} P_{\mathbf{Y}^{n}}\left(\mathbf{y}^{n}\right) \hspace{-0.5cm}\sum_{\substack{\mathbf{u'}^n:\mathbf{u'}^n \neq \mathbf{u}^n \\ (\mathbf{y}^n,\mathbf{u'}^n) \in \mathcal{T}_n^{(2)}}} e^{-nr} 
\end{align}
 From \eqref{T-2}, for $\left(\mathbf{y}^{n}, \mathbf{u'}^{n}\right) \in T_{n}^{(2)} $ we get $$P_{\mathbf{Y}^{n}}\left(\mathbf{y}^{n}\right) < P_{\mathbf{Y}^n \mid \mathbf{U}^n}\left(\mathbf{y}^n \mid \mathbf{u'}^n\right) e^{-n(r' -\epsilon)}, $$
which allows us to write
\begin{align}\notag
     \mathbb{P}\left(E_{12}\right) & \leq \sum_{\mathbf{u'}^n} \hspace{-0.1cm}\sum_{\mathbf{y}^{n}:(\mathbf{y}^n,\mathbf{u'}^n) \in \mathcal{T}_n^{(2)}} \hspace{-0.7cm} P_{\mathbf{Y}^n \mid \mathbf{U}^n}\left(\mathbf{y}^n \mid \mathbf{u'}^n\right) e^{-n(r + r' -\epsilon)} \\
     & \leq  e^{-n(r + r' - \overline{r}_0  -\epsilon)}
\end{align}
Therefore, from the condition $r \geq \overline{r}_0 - r' + \epsilon= \overline{I}(\mathbf{X} ; \mathbf{U}) - \underline{I}(\mathbf{Y} ; \mathbf{U}) + \epsilon$, we get that  $\mathbb{P}\left(E_{21}\right)\rightarrow 0$ as $n\rightarrow \infty$.

\textit{Type-II error $\beta_n$ :} A Type-II error occurs when the decoder declares $H_0$ although $H_1$ is the true hypothesis. The corresponding error events are:
 \begin{align}\notag
          E_{21} = &\left\lbrace\exists \Tilde{\mathbf{u}}^n \neq \mathbf{u}^n: \boldsymbol{B}(\Tilde{\mathbf{u}}^{n})=\boldsymbol{B}(\mathbf{u}^{n}) , \left(\overline{\mathbf{Y}}^n, \Tilde{\mathbf{u}}^{n}\right) \in T_{n}^{(2)}, \right.\\ \notag
       &\left. \text{ and} \left(\overline{\mathbf{Y}}^n, \Tilde{\mathbf{u}}^{n}\right) \in \mathcal{A}_n \right\},\\
      E_{22}=&\left\{(\overline{\mathbf{Y}}^{n},\mathbf{u}^{n})\in T_n^{(2)}, (\overline{\mathbf{Y}}^{n},\mathbf{u}^n)\in \mathcal{A}_n\right\}.
 \end{align}
 The first event $E_{21}$ is a debinning error and the second event $E_{22}$ is the testing error. 
 By the union bound, we get
 \begin{equation}
    \beta_n \leq \mathbb{P}\left(E_{21}\right)+\mathbb{P}\left(E_{22}\right).
    \label{beta}
\end{equation}
Since the marginal probability distribution $P_{\mathbf{Y}^n}$ does not depend on the hypothesis,  the probability $\mathbb{P}\left(E_{21}\right)$ can be expressed by following the same steps as for $\mathbb{P}\left(E_{12}\right)$. 
Given that $\overline{r}_{0} =  \overline{I}(\mathbf{X} ; \mathbf{U})$ and $r^{'} =  \underline{I}(\mathbf{Y} ; \mathbf{U})$ , we get
\begin{equation}
    \mathbb{P}\left(E_{21}\right) \leq e^{-n\left(r-\left(\overline{I}(\mathbf{X} ; \mathbf{U})-\underline{I}(\mathbf{Y} ; \mathbf{U})\right)-\varepsilon\right)}.
    \label{P_E21}
\end{equation}
Next, the probability $\mathbb{P}\left(E_{22}\right)$ can be expressed as
\begin{align}
 \mathbb{P}\left(E_{22}\right) &  \leq \sum_{(\mathbf{x}^n, \mathbf{y}^n)} P_{\overline{\mathbf{X}}^n \overline{\mathbf{Y}}^n}\left(\mathbf{x}^n , \mathbf{y}^n\right) \hspace{-0.65cm}\sum_{\substack{\mathbf{u}^n\in \llbracket 1,\text{M}_{1} \rrbracket, \\ (\mathbf{x}^n,\mathbf{u}^n) \in \mathcal{T}_n^{(1)}}} \hspace{-0.5cm} \mathbb{P} \left( (\mathbf{y}^{n},\mathbf{u}^{n} ) \in \mathcal{A}_{n}  \right) \notag \\
& \leq e^{n\overline{r}_0} \sum_{(\mathbf{x}^n, \mathbf{y}^n)} P_{\overline{\mathbf{X}}^n \overline{\mathbf{Y}}^n}\left(\mathbf{x}^n , \mathbf{y}^n\right)  \sum_{\substack{\mathbf{u}^n:  \\ (\mathbf{x}^n,\mathbf{u}^n) \in \mathcal{T}_n^{(1)} \\ (\mathbf{y}^n, \mathbf{u}^n) \in \mathcal{A}_{n}}} \hspace{-0.3cm} P_{\mathbf{U}^{n}}(\mathbf{u}^{n}) \notag
\end{align}
Since $(\mathbf{x}^{n}, \mathbf{u}^{n})\in \mathcal{T}_n^{(1)}$,
$$
P_{\mathbf{U}^{n}}(\mathbf{u}^{n}) < P_{\mathbf{U}^{n}\mid \mathbf{X}^{n}} (\mathbf{u}^{n} \mid \mathbf{x}^{n}) e^{-n(\underline{r}_0-\epsilon)}.
$$
In addition, the conditional distributions $P_{\mathbf{U}^{n}\mid \mathbf{X}^{n}}$ and $P_{\overline{\mathbf{U}}^{n}\mid \overline{\mathbf{X}}^{n}}$ are the same, and the Markov chain $\mathbf{U}^n \rightarrow \mathbf{X}^n \rightarrow \mathbf{Y}^n$ is satisfied. Thus, $P_{\mathbf{U}^{n}\mid \mathbf{X}^{n}} = P_{\mathbf{\overline{U}}^{n}\mid \mathbf{\overline{X}}^{n},\overline{Y}^n}$, and
\begin{align}
  \mathbb{P}\left(E_{22}\right) & \leq e^{n(\overline{r}_0 - \underline{r_0} + \epsilon)} \sum_{\mathbf{u}^n:  (\mathbf{y}^n, \mathbf{u}^n) \in \mathcal{A}_{n}} P_{\overline{\mathbf{U}}^n \overline{\mathbf{Y}}^n}\left(\mathbf{u}^n , \mathbf{y}^n\right).
  \label{theta_det1}
\end{align}
For $\left(\mathbf{y}^{n},\mathbf{u}^{n}\right)  \in \mathcal{A}_{n}$, we have 
\begin{equation}
    P_{\overline{\mathbf{U}}^n \overline{\mathbf{Y}}^n}\left(\mathbf{u}^n , \mathbf{y}^n\right) < P_{\mathbf{U}^n \mathbf{Y}^n}\left(\mathbf{u}^n , \mathbf{y}^n\right) e^{-n(S-\epsilon)}.
    \label{theta_det2}
\end{equation}
Combining this with~\eqref{theta_det1} gives that
\begin{equation}
\mathbb{P}\left(E_{22}\right)  \leq  e^{-n( \underline{r}_0 - \overline{r_0}  + S-2\epsilon)} 
    \label{P_E22}
\end{equation}
Now, substituting \eqref{P_E21} and \eqref{P_E22} into \eqref{beta},  with $S=\underline{D}\left(P_{\mathbf{U}\mathbf{Y}} \| P_{\overline{\mathbf{U}}\overline{\mathbf{Y}}}\right)$, the Type-II error is upper-bounded as
\begin{align}\notag
         \beta_n  \leq & e^{-n\left(r-\left(\overline{I}(\mathbf{X} ; \mathbf{U})-\underline{I}(\mathbf{Y} ; \mathbf{U})\right)-\epsilon\right)} \\ & +e^{-n\left(  \underline{I}(\mathbf{X};\mathbf{U}) - \overline{I}(\mathbf{X};\mathbf{U})+  \underline{D}\left(P_{\mathbf{U}\mathbf{Y}} \| P_{\overline{\mathbf{U}}\overline{\mathbf{Y}}}\right)-2\epsilon\right)}.
    \label{beta_bin_test}
\end{align}
Finally, from the definition of the error exponent $\theta$ given by~\eqref{betha_const}, we show that \eqref{th1} is achievable, which proves Theorem \ref{theorem_1}.

\section{Examples} \label{examples}
We now apply Theorem \ref{theorem_1} to some source models of interest.
\subsection{i.i.d. sources}
We first consider i.i.d. sources $\mathbf{X}$ and $\mathbf{Y}$ in 
 order to check the consistency of Theorem~\ref{theorem_1} with respect to existing results in the literature. 
 We here assume that the pairs $(\mathbf{U}^{n}, \mathbf{X}^{n})$ and $(\mathbf{U}^{n}, \mathbf{Y}^{n})$ are i.i.d. according to the joint distributions $P_{UX}$ and $P_{UY}$, respectively. In this case, according to~\cite[Page 18]{Han1998}, the spectral terms involved in~\eqref{th1} are equal to their conventional counterparts, and hence~\eqref{th1} 
  becomes 
\begin{equation}\notag
    \theta \leq  \operatorname{min} \left\{ r-\left(I(X ;U)-I(Y ;U)\right),  D\left(P_{UY} \| P_{\overline{U}\overline{Y}}\right) \right\}.
 \label{theta_iid}
 \end{equation}
As expected, we find that our error exponent is consistent with that shown in \cite{Katz2015} for the i.i.d. case. On the other hand, our error-exponent differs from the Shimokawa-Han-Amari error-exponent obtained in~\cite{shimokawa1994error}. This comes from the fact that our achievability scheme performs two steps at the decoder: debinning and testing, while the scheme of~\cite{shimokawa1994error} performs only one step. 


\subsection{Stationary and ergodic sources}
We then consider that the sources are stationary and ergodic, but not necessarily i.i.d.
\begin{Proposition}
If the sources $\mathbf{X}^n$ and $\mathbf{Y}^n$ are stationary and ergodic under both $H_0$ and $H_1$, the error exponent \eqref{th1} becomes :
\begin{equation}
\begin{split}
    \theta \leq  \operatorname{min} \left\{ \lim _{n \rightarrow \infty} r - \left[\frac{1}{n}h\left(\mathbf{U}^n \mid \mathbf{Y}^n\right)- \frac{1}{n}h\left(\mathbf{U}^n \mid \mathbf{X}^n\right)\right], \right.\\ 
     \left.\lim _{n \rightarrow \infty} \frac{1}{n} D \left(P_{\mathbf{U}^{n}\mathbf{Y}^{n}} \| P_{\overline{\mathbf{U}}^{n}\overline{\mathbf{Y}}^{n}}\right)\right\} . 
\end{split}
    \label{theta_stat_erg} 
\end{equation}
\label{prop_stat_erg}
\end{Proposition}
\vspace{-0.2cm}
This proposition is due to the \textit{strong converse property} \cite[Page 48-49]{Han1998}.  
\subsection{Stationary and ergodic Gaussian sources}
Let $\mathbf{X}$ and $\mathbf{Y}$ be two stationary and ergodic sources distributed according to Gaussian distributions $\mathcal{N}\left(\mathbf{\mu}_{\mathbf{X}}, \mathbf{K}_{\mathbf{X}}\right) \text{and } \mathcal{N}\left(\mathbf{\mu}_{\mathbf{Y}}, \mathbf{K}_{\mathbf{Y}}\right)$, with covariance matrices $\mathbf{K}_{\mathbf{X}}$ and $\mathbf{K}_{\mathbf{Y}}$, respectively. 
The two hypotheses are formulated as 
\begin{equation}
   H_0 :\left(\begin{array}{l}
\mathbf{X}^{n} \\
\mathbf{Y}^{n}
\end{array}\right) \sim \mathcal{N}(\mathbf{\mu}_{\mathbf{XY}}, \mathbf{K}),
\label{h0_vectx}
\end{equation}
\begin{equation}
   H_1 :\left(\begin{array}{l}
\mathbf{X}^{n} \\
\mathbf{Y}^{n}
\end{array}\right) \sim \mathcal{N}(\overline{\mathbf{\mu}}_{\mathbf{XY}}, \overline{\mathbf{K}}).
\label{h1_vectx}
\end{equation}
In the expressions \eqref{h0_vectx} and \eqref{h1_vectx}, $\mathbf{\mu}_{\mathbf{XY}}$ is defined as a block vector $\left[\mathbf{\mu}_{\mathbf{X}},\mathbf{\mu}_{\mathbf{Y}} \right]^{T}$. 
In addition, $\mathbf{K}$ and $\overline{\mathbf{K}}$ are the joint covariance matrices of $\mathbf{X}$ and $\mathbf{Y}$ defined as 
\begin{equation}
    \mathbf{K}=\left[\begin{array}{ll}
\mathbf{K}_{\mathbf{X}} & \mathbf{K}_{\mathbf{X Y}} \\
\mathbf{K}_{\mathbf{Y X}} & \mathbf{K}_{\mathbf{Y}}
\end{array}\right], 
    \overline{\mathbf{K}}=\left[\begin{array}{ll}
\mathbf{K}_{\mathbf{X}} & \overline{\mathbf{K}}_{\mathbf{X Y}} \\
\overline{\mathbf{K}}_{\mathbf{Y X}} & \mathbf{K}_{\mathbf{Y}}
\end{array}\right] ,
\label{K1_block_xy}
\end{equation}
Although not explicit in our notation, we here consider that the vectors $\mathbf{\mu}_{\mathbf{X}}$ and $\mathbf{\mu}_{\mathbf{Y}}$ are of length $n$, and that the covariance matrices $\mathbf{K}$ and $\overline{\mathbf{K}}$ are of size $2n\times 2n$, where $n$ will tend to infinity in the subsequent analysis.
We assume that all the matrices $\mathbf{K}_{\mathbf{X}}$, $\mathbf{K}_{\mathbf{Y}}$, $\overline{\mathbf{K}}_\mathbf{Y}$, $\mathbf{K}_{\mathbf{X Y}}$, and $\overline{\mathbf{K}}_{\mathbf{XY}}$ are positive-definite. We also denote the conditional covariance matrix of $\mathbf{X}^n$ given $\mathbf{Y}^n$ by
\begin{equation}
    \mathbf{K}_{\mathbf{X \mid Y}} =\mathbf{K}_{\mathbf{X}}-\mathbf{K}_{\mathbf{XY}}\mathbf{K}_{\mathbf{Y}}^{-1} \mathbf{K}_{\mathbf{XY}}.
    \label{K_x/y}
\end{equation}
The eigenvalues of $\mathbf{K}_{\mathbf{X \mid Y}}$ are  further denoted by $\lambda_i^{(X \mid Y)}$.\\
\begin{Proposition} If the sources $\mathbf{X}$ and $\mathbf{Y}$ are Gaussian, stationary, and ergodic, under both $H_0$ and $H_1$, the terms in $\eqref{theta_stat_erg}$ reduce to 
\begin{align}
    \lim _{n \rightarrow \infty} \frac{1}{n}h\left(\mathbf{U}^n \mid    \mathbf{Y}^n\right)- \lim _{n \rightarrow \infty} \frac{1}{n}h\left(\mathbf{U}^n \mid \mathbf{X}^n\right) = \notag\\
    \lim _{n \rightarrow \infty} \frac{1}{2n} \sum_{i=1}^n \log \frac{\lambda_i^{(X \mid Y)}+\kappa}{\kappa},
     \label{entropy_rate} 
\end{align} 
and
\begin{align}
& \lim _{n \rightarrow \infty} \frac{1}{n} D \left(P_{\mathbf{U}^{n}\mathbf{Y}^{n}} \| P_{\overline{\mathbf{U}}^{n}\overline{\mathbf{Y}}^{n}}\right\}  = \lim _{n \rightarrow \infty} \frac{1}{2n} \left[ \log \frac{\left|\overline{\mathbf{\Sigma}}\right|}{\left|\mathbf{\Sigma}\right|} - 2n\  + \right. \notag \\
& \left.\left(\overline{\mathbf{\mu}}_{\mathbf{UY}}-\mathbf{\mu}_{\mathbf{UY}}\right)^{T}\overline{\mathbf{\Sigma}}^{-1} \left(\overline{\mathbf{\mu}}_{\mathbf{UY}}-\mathbf{\mu}_{\mathbf{UY}}\right)+\operatorname{tr}\left\{\overline{\mathbf{\Sigma}}^{-1} \mathbf{\Sigma}\right\} \right],
\label{divergence-rate}
\end{align}
where $\mathbf{\Sigma}$ and $\overline{\mathbf{\Sigma}}$ are the joint covariance matrices of $\mathbf{U}$ and $\mathbf{Y}$ under $H_0$ and $H_1$, respectively.
 \label{prop_error_stat_ergo_gauss}   
\end{Proposition}
The terms given by \eqref{entropy_rate} and \eqref{divergence-rate} are obtained by considering that the source $\mathbf{U}$ is Gaussian such that 
  $\mathbf{U} = \mathbf{X}+\mathbf{Z}$,
where $\mathbf{Z} \sim \mathcal{N} \left(0,\kappa\mathbf{I}_n\right)$ is independent of $\mathbf{X}$, and $\mathbf{I}_n$ is the identity matrix of dimension $n\times n$. The covariance matrices $\mathbf{\Sigma}$ and $\overline{\mathbf{\Sigma}}$ are then defined as 
\begin{equation}
    \mathbf{\Sigma}=\left[\begin{array}{ll}
\mathbf{K}_{\mathbf{U}} & \mathbf{K}_{\mathbf{U Y}} \\
\mathbf{K}_{\mathbf{Y U}} & \mathrm{K}_{\mathbf{Y}}
\end{array}\right],
    \overline{\mathbf{\Sigma}}=\left[\begin{array}{ll}
\mathbf{K}_{\mathbf{U}} & \overline{\mathbf{K}}_{\mathbf{U Y}} \\
\overline{\mathbf{K}}_{\mathbf{Y U}} & \mathrm{K}_{\mathbf{Y}}
\end{array}\right] .
\label{sigma1_block_uy}
\end{equation}
We now consider the case where the pair $(\mathbf{U},\mathbf{Y)}$ has different covariance matrices, $\mathbf{\Sigma}$ under $H_0$ and $\overline{\mathbf{\Sigma}}$ under $H_1$. We also assume that all the Gaussian vectors are zero-centered. We then define $H_0$ and $H_1$ as
\begin{equation}
   H_0 :\left(\begin{array}{l}
\mathbf{X}^{n} \\
\mathbf{Y}^{n}
\end{array}\right) \sim \mathcal{N}(\mathbf{0}, \mathbf{K}),
\label{h0_cov}
\end{equation}
\begin{equation}
   H_1 :\left(\begin{array}{l}
\mathbf{X}^{n} \\
\mathbf{Y}^{n}
\end{array}\right) \sim \mathcal{N}(\mathbf{0}, \overline{\mathbf{K}}).
\label{h1_cov}
\end{equation}
In this case, it can be shown that the expression \eqref{entropy_rate} remains the same, while the expression \eqref{divergence-rate} reduces to
\begin{equation}
\begin{split}
    \lim _{n \rightarrow \infty} \frac{1}{n} D \left(P_{\mathbf{U}^{n}\mathbf{Y}^{n}} \| P_{\overline{\mathbf{U}}^{n}\overline{\mathbf{Y}}^{n}}\right\}  = \lim _{n \rightarrow \infty} \frac{1}{2n}\left[ \log \frac{\left|\overline{\mathbf{\Sigma}}\right|}{\left|\mathbf{\Sigma}\right|} \right. \\ 
   \left. - 2n + \operatorname{tr}\left\{\overline{\mathbf{\Sigma}}^{-1} \mathbf{\Sigma}\right\} \right].
\end{split}
\label{divergence-rate-covariance}
\end{equation}
Note that the matrices $\mathbf{\Sigma}$ and $\overline{\mathbf{\Sigma}}$ are of length $2n \times 2n$. Therefore, to specify the previous result to some specific Gaussian sources, one needs to study the convergence of the determinants $\left|\mathbf{\Sigma}\right|$ and $\left|\overline{\mathbf{\Sigma}}\right|$, and also of the trace $\operatorname{tr}\left\{\overline{\mathbf{\Sigma}}^{-1} \mathbf{\Sigma}\right\}$. 
\section{Conclusion} 
\label{conclusion}
This work studies the DHT problem for general non-i.i.d., non-stationary, and non-ergodic sources. It uses an information spectrum approach to provide a general formula for the achievable Type-II error exponent. The achievability proof is based on a quantize-and-binning coding scheme and the derived error exponent boils down to a trade-off between a binning error and a decision error, as already observed for i.i.d. sources. Future works will include considering a hidden Markov correlation model between the source $\mathbf{X}$ and the side information $\mathbf{Y}$, as well as designing practical coding schemes for DHT. Application to synchronism identification in spread spectrum signal detectors~\cite{arribas2013antenna} will also be considered.





\bibliographystyle{IEEEtran}
\bibliography{papers_done}

\end{document}